\documentclass[12pt, oneside]{amsart}   	% use "amsart" instead of "article" for AMSLaTeX format
\usepackage{geometry}                		% See geometry.pdf to learn the layout options. There are lots.
\usepackage{graphicx}				% Use pdf, png, jpg, or epsß with pdflatex; use eps in DVI mode
\usepackage{cite}
			
\usepackage{amssymb}

\makeatletter
\def\blfootnote{\gdef\@thefnmark{}\@footnotetext}
\makeatother

\title{Raindrop impact on sand: a dynamic explanation of crater morphologies}
\author{Song-Chuan Zhao$^\#$, Rianne de Jong, \and Devaraj van der Meer}

\begin{document}

\begin{abstract}
As a droplet impacts on a granular substrate, both the intruder and the target deform, during which 
the liquid may penetrate into the substrate. {These three aspects together distinguish} it from other 
impact phenomena in the literature. We perform high-speed, double-laser profilometry measurements and disentangle the dynamics into three aspects: the deformation of the substrate during the impact, the maximum spreading diameter of the droplet, and the penetration of the liquid into the substrate. By systematically varying the impact speed and the packing fraction of the substrate, (i) the substrate deformation indicates a critical packing fraction $\phi^*\approx 0.585$; (ii) the maximum droplet spreading diameter is found to scale with a Weber number corrected by the substrate deformation; and (iii) a model about the liquid penetration is established and is used to explain the observed crater morphology transition.
\end{abstract}

\maketitle

\blfootnote{\textit{${\#}$~Physics of Fluids Group, Faculty of Science and Technology, University of Twente, PO Box 217, 7500 AE Enschede, The Netherlands; E-mail: szhao@utwente.nl}}

\begin{figure}
\centering
\includegraphics[width=8.3cm]{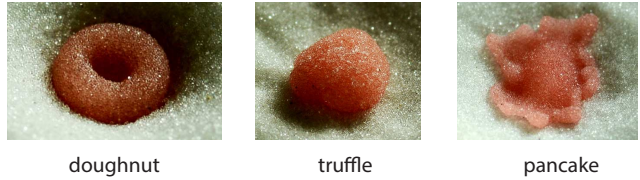} 
\caption{\label{f.morph} Various crater morphologies observed in the experiments are categorized as: doughnut, truffle, and pancake.}
\end{figure}

Liquid droplet impact on a granular layer is very common in nature, industry, and agriculture{, ranging} from raindrop{s} falling {onto desert or soil} %on earth or desert 
to granulation in the production process of many pharmaceuticals. In spite of its commonness, the physical mechanism{s} involved in the impact of a droplet on sand did not attract much attention {until} recently\cite{Katsuragi2010,Marston2010,Katsuragi2011,Delon2011,Nefzaoui2012,Marston2012,Hamlett2013,Long2014,Zhao2014,Emady2011,Emady2013,Emady2013a},  and the underlying physics is still largely unexplored. In contrast, droplet impact on a solid surface or a liquid pool {has} %have 
been studied extensively\cite{Rein1993,CLANET2004,Eggers2010}. However, a granular substrate can act both solid-like and fluid-like\cite{Jaeger1996} {and m}any experiments {have} been conducted to reveal the response of a granular packing to a solid object impact\cite{Amato1998,Uehara2003,Lohse2004,Katsuragi2007,Umbanhowar2010,Gravish2010,Ruiz-Suarez2013}, where the intruder does not deform. {D}roplet impact on a granular substrate adds new challenges to the above: {F}irst, both the intruder and the target, not merely one of the two, deform during impact; second, the liquid composing the droplet may penetrate into the substrate during the impact and may, in the end, completely merge with the grains. These complex interactions between the droplet intruder and the granular target create various crater morpholog{ies} as reported in the literature\cite{Katsuragi2010,Delon2011,Marston2010,Emady2013a,Zhao2014} {[see }Fig.~\ref{f.morph} for examples]
. An appealing and natural question is {by what} mechanism {craters are formed and how this leads to the observed rich morphological variation}. This is the main focus of the present work. Quantitative dynamic details, \textit{e.g.}, {on} the deformation of droplet and substrate in response to the impact, are necessary to gain insight about this issue. Previous dynamic measurements only regard the droplet spreading\cite{Delon2011,Marston2010,Nefzaoui2012} or individual splashing grains\cite{Long2014}, and a systematical study {of}
the effect of the packing fraction of the substrate is still missing. In this paper, we perform dynamic measurements with high-speed laser profilometry and study the dependence of the dynamics on impact speed and packing fraction of the granular substrate. Following the discussion about the results of the substrate deformation and the droplet deformation and the underlying physics, a {quantitative} model is established %with the quantified dynamics to 
{which explains} the %various 
{observed} crater morphologies in the end. 

\section{Experimental methods}

\begin{figure}[h]
\centering
\includegraphics[width=8.3cm]{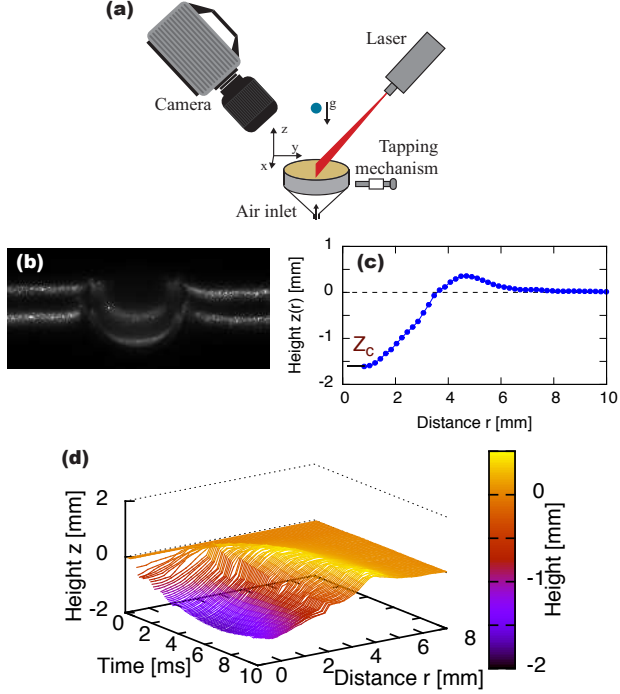}
\caption{\label{f.setup} \textbf{(a)} Setup sketch. %$x$, $y$ denote the horizontal plane. 
The height, $z(x,y)$, is measured as a function of coordinates on {the horizontal} $xy$ plane with laser sheets and a high speed camera. \textbf{(b)} An image taken during the impact, where the droplet is at its maximum deformation. Two laser lines are used {for high-speed profilometry, where t}he deflection of the laser lines indicates the deformation of the target surface. \textbf{(c)} The laser line deflection{ in }%the image in 
(b) is used to reconstruct the {height}, $z$, as a function of the distance to the impact center, $r$. The dashed line indicates the initial surface height, $z=0$ {and t}he center depth is labeled{ }$Z_c$. \textbf{(d)} Time evolution $z(r,t)$ of the crater depth (color coded).}
\end{figure}

A water droplet with diameter $D_0=2.8\, \mathrm{mm}$ acts as the intruder. The water droplet {is} pinched off from a needle at rest {and dropped from} a certain height. The droplet is accelerated by gravity and impacts on the horizontal surface perpendicularly. The impact speed, $U_0$, is computed from a calibrated height-speed profile. {Rain consists of droplets with a maximum diameter 6 mm}\cite{Villermaux2009} {falling at their terminal velocity. For a droplet diameter of 2.8 mm, the terminal velocity is about 7 m/s}. {In this study however, the impact speed is one of the important control parameters and is varied from 1.35 m/s to 4.13 m/s.} The target granular packing consists of polydisperse soda lime glass beads with diameter $d_g=70-110\, \mathrm{\mu m}$ and specific density {$\rho_g=2500\, \mathrm{kg/m^3}$. While there are studies considering the effect of the wettability of grains}\cite{Marston2012,Hamlett2013}, {the wettability is not varied in this work, where hydrophilic grains are used.} Grains are put into {an} oven at $90\,^{\circ}\mathrm{C}$ for at least half hour before experiments, {after which we let them cool down to room temperature}. Various packing fractions, $\phi_0$, are prepared by air fluidization and a tapping device [Fig.~\ref{f.setup}a]. The initial packing fraction $\phi_0$ is computed according to the surface height relative to the container edge before impact. {During the impact, t}he height of the substrate surface, $z(x,y)$, is measured by {high-speed} laser profilometry with a typical depth resolution of 0.1 mm/pixel. The impact process is captured by a high speed camera (Photron SA-X2) with a frame rate {of} 10,000~Hz, {of which an} experimental image is shown in Fig.~\ref{f.setup}b. For dynamic analysis, the surface height before impact is taken as the reference $z=0$. {We use t}wo laser lines {for the} profilometry , and, as the surface deforms, the lines are deflected. We translate the deflection into the height function, $z(r)$, where $r$ is the distance to the impact center in the horizontal plane, \textit{i.e.}, $r=\sqrt{(x-x_c)^2+(y-y_c)^2}$, where $x_c$, $y_c$ are the coordinates of the impact center. Assuming {axisymmetry}, the impact center is located {by determining} when the deflections reconstructed from individuals laser lines match. Henceforth, the height function, $z(r)$, used for further analysis is {always} averaged over {the} two laser lines. An example of $z(r)$ and its time evolution are illustrated in Fig.~\ref{f.setup}(c) and (d).

\section{Substrate deformation}
%\label{sec.substrate}
\begin{figure}
\centering
\includegraphics[width=8.7cm]{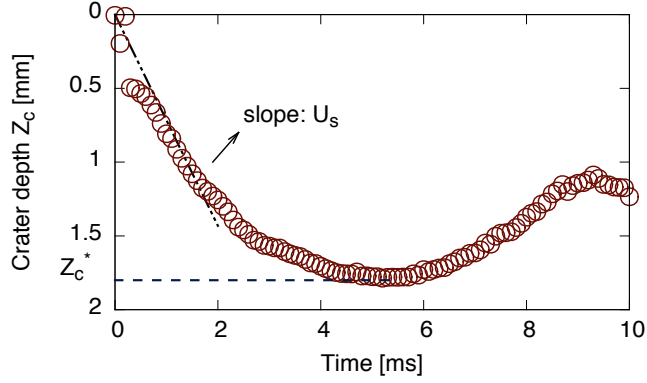}

\caption{\label{f.depth}Crater depth versus time. The maximum crater depth is denoted as as $Z_c^*$. A straight line is fitted on the data points of $t\leq (D_0+2Z_c^*)/U_0$. The slope of {this fit}, $U_s$, is used to indicate the speed of the deformation. {Here, $U_0=3.19$ m/s, and $\phi_0 = 0.569$.}}

\end{figure}

We use the depth of the crater center, $Z_c(t)$ as defined in Fig.~\ref{f.setup}(c), to characterize the development of the crater. A typical temporal evolution of the crater depth is shown in Fig.~\ref{f.depth}. In the beginning the crater becomes deeper with time. At a certain moment, the crater reaches its maximum depth $Z_c^*$, {after which} the droplet contracts and {transports} grains mixed with it to{wards} the crater center. An avalanche subsequently occurs at a longer time scale. Both of these effects tend to {decrease} 
$Z_c$, {\textit{i.e.}, make the crater shallower}.  Here, we focus only on the early stage of $Z_c(t)$ and quantify its evolution {by measuring} %with
two quantities: the {initial} speed of the deformation, $U_s$, and the maximum crater depth, $Z_c^*$\cite{note1}. The behavior of these two quantities is essential to understand not only the response of the granular substrate, but also the deformation of the droplet and the formation of various crater shapes.

\subsection{Deformation speed $U_s$}
%\label{subsec.speed}

\begin{figure}
\centering
\includegraphics[width=8.3cm]{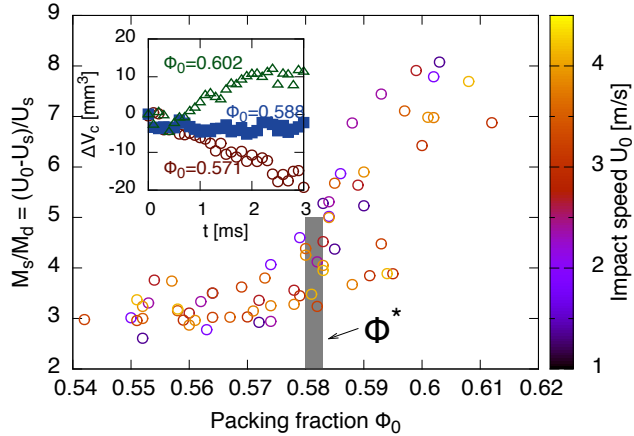}

\caption{\label{f.dilation_mech} Main plot: {{The ratio of the impact velocity $U_0$ and the deformation speed of the substrate $U_s$} as a function of packing fraction $\phi_0$.} The Y-axis, $(U_0-U_s)/U_s$, indicates the amount of grains involved in the deformation {$M_s$ normalized by the droplet mass $M_d$} (see text), where $U_s$ is the speed of the substrate deformation [Fig.~\ref{f.depth}]. Inset: the volume change of the substrate, $\Delta V_c$, for experiments with the same impact speed $U_0=3.75~\mathrm{m/s}$, but three different packing fraction{s}, $\phi_0=0.571,0.588,0.602$, from bottom to top. }

\end{figure}

The {initial} deformation speed $U_s$ is {determined as} %evaluated by 
the slope of a linear fit of $Z_c(t)$ {within a time duration $t\leq t_{imp}$, where $t_{imp}=(D_0+2Z_c^*)/U_0$ denotes the impact time scale.} [Fig.~\ref{f.depth}]. {The ratio of the impact velocity and this} deformation speed  
{is} plotted in Fig.~\ref{f.dilation_mech} for all experiments. {This ratio} is {largely} independent of impact speed which implies that {it} is an inherent property of {the} granular substrate. On the other hand, the dependence on {the packing fraction} $\phi_0$ %illustrates 
{indicates} a transition around $\phi^*\approx 0.585$. \footnote{{Although the measurement of $U_s$ is merely the first approximation of the deformation speed, we confirm that $\phi^*$ is not altered by reducing the fitting region to half its size.}} Here, we introduce a simple scenario to elucidate this transition. Upon impact the droplet {--} with a mass $M_d$ {--} accelerates a certain amount of grains {in the vertical direction} to {the initial deformation speed} $U_s$, while  {by vertical momentum conservation} the momentum lost by the droplet is proportional to $M_d(U_0-U_s)$. {Thus, t}he mass of grains accelerated by the droplet is  $(U_0-U_s)/U_s$ times larger than the mass of the droplet. {This ratio,} plotted in Fig.~\ref{f.dilation_mech}, {therefore} indicates the amount of substrate material involved in the deformation dynamics and the yield stress of the substrate, \textit{i.e.}, the larger this ratio is, the harder the substrate is. With this interpretation, the observed transition at $\phi^*$ in Fig.~\ref{f.dilation_mech} is reminiscent of the penetration force transition which points to the dilatancy transition\cite{Schroter2007}.  

The dilatancy transition expresses a peculiar phenomena that shearing a granular packing above a threshold packing fraction results in dilation {(\textit{i.e.}, expansion along the perpendicular directions)}, whereas a loose packing {is} just compactified under {such a perturbation}\cite{Thompson1991}. Though the word `dilation' describes a change of the volume, previous studies have shown that the dilatancy transition is accompanied by a force response transition\cite{Schroter2007,Gravish2010,Umbanhowar2010,Metayer2011}.  As explained above, this force transition feature is already encapsulated by the factor $(U_0-U_s)/U_s$ representing momentum transfer, and the volume change of the substrate is supposed to give a transition at $\phi^*$ as well. {We compute t}he volume change 
{as} the integral, $\Delta V_c=\pi\int z(r)dr^2$, for each frame. The result is plotted in the inset of Fig.~\ref{f.dilation_mech} {for three} experiments with different $\phi_0$ but the same impact speed $U_0$. The volume change is negative {(\textit{i.e.}, the substrate compactifies)} for {the loosest} substrate and indeed increases to positive {(indicating dilation)} with increasing $\phi_0$. Based on the above evidence we define a critical packing fraction $\phi^*\approx 0.585$.\footnote{{Note that these measurements pertain to the initial deformation of the bed, \textit{i.e.}, before liquid-grain mixing becomes important.} The dilation and its transition reported here are inherent properties of granular packings, independent of the intruder properties. This is different to the dilation measured from the final crater volume\cite{Katsuragi2011}, where the residue of the liquid-grain mixture plays a very important role.} The grains underneath the droplet intruder are forced to rearrange to $\phi^*$ during the impact whatever {the} initial packing fraction $\phi_0$ is. {This value will be used to model the mixing between liquid and grains}.

\subsection{Maximum crater depth $Z_c^*$}
\begin{figure}[h]
\centering
  \includegraphics[width=8.3cm]{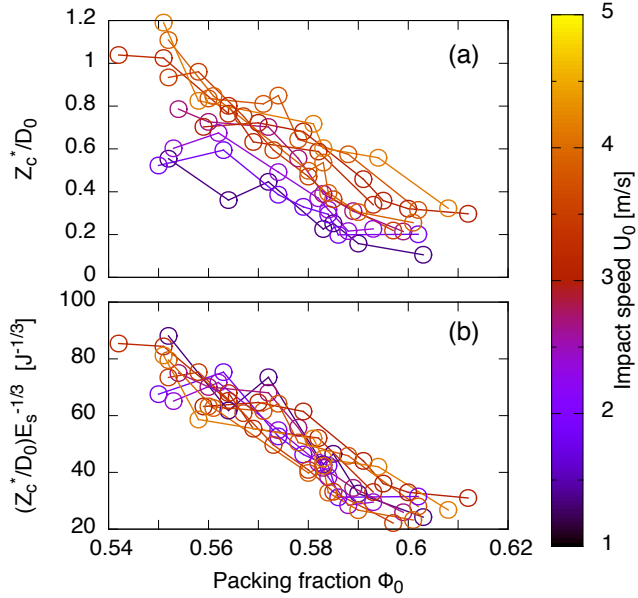}
  \caption{\label{f.zmax} \textbf{(a)}: Maximum crater depth $Z_c^*$ {(non-dimensionalized with the droplet radius $D_0$)} versus packing fraction $\phi_0$. {The i}mpact velocity, $U_0$, is color coded. The dependence on $\phi_0$ is stronger than that on $U_0$. \textbf{(b)}: Maximum crater depth {scaled} with the {cubic root of the} energy transferred into the substrate, $E_s${, again versus $\phi_0$}. See text and Eq.~\ref{eq.energy_ratio} for details.}
  
\end{figure}

In contrast to Fig.~\ref{f.dilation_mech}, the maximum crater depth, $Z_c^*$ plotted in Fig.~\ref{f.zmax}a, depends on both {the} initial packing fraction, $\phi_0$, and the impact speed of the droplet, $U_0$. 
It has been shown above that a dense substrate is more difficult to deform, therefore it {should come as no surprise} to see that the maximum crater depth $Z_c^*$ decreases with $\phi_0$. The {increase} of  $Z_c^*$ with the impact speed $U_0$ is also anticipated: a higher impact energy generates a deeper crater.

Studies on solid intruder impact on a granular substrate\cite{Amato1998,Uehara2003,Katsuragi2007} can %further 
help to {further} understand the impact speed dependence{.} %here. 
For {a} %the 
crater created by a solid object, two scaling laws of its depth with impact energy are suggested based on how the kinetic energy, $E_k$, is dissipated\cite{Amato1998}. {Assuming that} plastic deformation is the most dissipating process{,} %then 
the crater volume scales with the kinetic energy, which yields $Z_c^*\sim E_k^{1/3}$\cite{Uehara2003,Katsuragi2007}. On the other hand, if material ejection absorbs most of the impact energy, then the converted gravitational energy %calls for 
{leads to} $Z_c^*\sim E_k^{1/4}$. The volume change in Fig.~\ref{f.dilation_mech} implies strong plastic deformations during the impact.  However, in the case of droplet impact, not only the substrate but also the intruder dissipates the impact energy via deformations. To apply the above arguments, we {therefore} first need to decide {upon} the distribution of $E_k$ between the droplet and the granular packing.

The droplet experiences a deceleration force. This force {does work on both the droplet and the granular target along the total displacement, $\frac{1}{2}D_0+Z_c^*$. This work transforms the impact energy $E_k=M_dU_0^2/2$ into other energy forms, such as surface energy of the droplet and dissipation inside the substrate}. Out of the total displacement, $\frac{1}{2}D_0$ is the contribution of the droplet deformation, while $Z_c^*$ is that of the substrate deformation. Such a force, $E_k/(\frac{1}{2}D_0+Z_C^*)$, indicates the {average} interaction between droplet and the substrate{, from which we can estimate} {the energy absorbed by the droplet} $E_d$ and {that absorbed by the substrate} $E_s$ {as} %equals to 
the work done by {this} interaction force to deform the droplet and substrate respectively:
\begin{equation}
E_d = \frac{D_0}{D_0+2Z_c^*}E_k;~ E_s = \frac{Z_c^*}{D_0/2+Z_c^*}E_k
\label{eq.energy_ratio}
\end{equation}
{This energy distribution is estimated using the average force between the intruder and the target. While it has been shown that the impact force between a droplet and a solid surface is time dependent}\cite{Soto2014, Eggers2010}, {the subsequent analysis justifies} Eq.~\ref{eq.energy_ratio} {by hindsight.}  

With measured $Z_c^*$ and $E_k$, the energy distributed into the substrate $E_s$ can be computed for all experiments. We find that $Z_c^*\propto E_s^{1/3}$ is the better relation {to collapse} the dependence on $U_0$ [Fig.~\ref{f.zmax}b]. {For a given $\phi_0$, a power law fit gives $Z_c^*\propto E_s^{\alpha}$ with $\alpha = 0.33 \pm 0.04$. This scaling} suggests plastic deformation {as the main cause of energy dissipation, which stands to reason}. We will show that the kinetic energy distribution in Eq.~\ref{eq.energy_ratio} has essential consequences to the droplet deformation and the crater morphology as well.

\section{Maximum droplet diameter $D_d^*$}
%\label{sec.droplet}
In contrast to solid object impact, the droplet deforms during impact\cite{Delon2011,Marston2010,Nefzaoui2012}. This droplet deformation %{is leading for} %leads
%the crater {open} at the early stage of the impact, and {in addition} %also 
changes the contact area between liquid and grains, which {will turn out to play an important role} in determining the crater morphology. In this section we focus on the maximum spreading of the droplet.
  
For most of our experiments, {the} droplet contracts after spreading, which implies that surface tension is the dominant stopping force of droplet deformation. In this regime, {the} Weber number {$\textrm{We} = \rho_lD_0U_0^2/\sigma$} is the {expected} relevant dimensionless parameter measuring the relative importance of the kinetic energy of the impact droplet and the surface energy. {Here, }$\rho_l$ is the %specific 
density of the liquid {and} $\sigma$ is its surface tension. Previous studies about droplet impact on a solid substrate have proposed two models for maximum %droplet 
diameter, $D_d^*${, that is reached by the droplet}. One is based on energy conservation argument where the kinetic energy is converted to the surface energy. This model results in a scaling $D_d^*/D_0\propto \textrm{We}^{1/2}$\cite{Bennett1993}. The other more recent model balances the inertial force with the surface tension and suggests $D_d^*/D_0\propto \textrm{We}^{1/4}$\cite{CLANET2004}. In the latter case significant amount of the impact energy is dissipated into internal degrees of the droplet.

\begin{figure}
\centering
  \includegraphics[width=8.3cm]{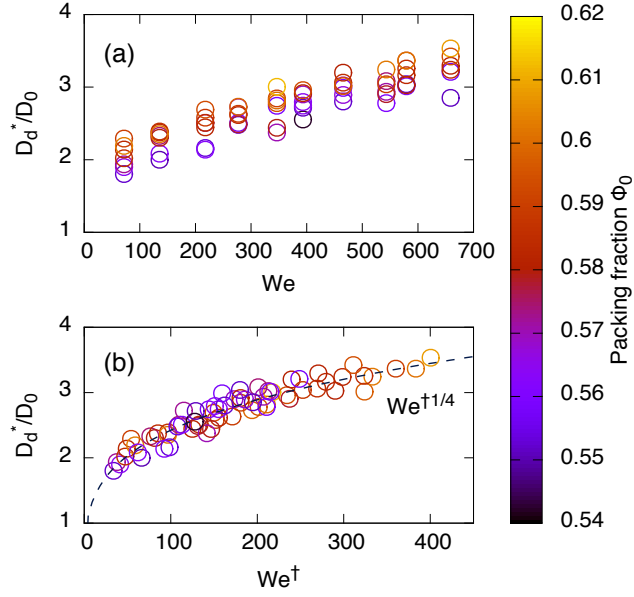}
  \caption{\label{f.dmax} Maximum droplet deformation $D_d^*$ normalized with the initial droplet diameter $D_0$ versus Weber number {$\textit{We}=\rho_lD_0U_0^2/\sigma$} in (a) and the effective Weber {number} ${{\textit{We}}^{\dagger}} = D_0/(D_0+2Z_c^*) \textit{We}$ in (b), where $Z_c^*$ is the maximum crater depth [Fig.~\ref{f.depth}]. Packing fraction is color coded. {The dashed line in (b) indicates $D_d^*/D_0 \propto {{\textit{We}}^{\dagger}}^{1/4}$}.}
  
\end{figure}

{In Fig.~}\ref{f.dmax}{a, t}he maximum droplet %deformations 
{diameter} in our experiments {is} plotted {against the} Weber number $\textit{We}$ {for different} packing fractions. {We observe that} data with the same impact speed $U_0$ are scattered {due to the variation in} packing fraction{, where a} denser packing typically generates {a} larger droplet deformation. This can be understood {from} %with 
the dependence of the substrate deformation on the initial packing fraction [Fig.~\ref{f.zmax}] together with Eq.~\ref{eq.energy_ratio}. It has been shown that a denser packing deforms less, thus more energy is transferred into the droplet resulting in a larger spreading. In the energy conservation model\cite{Bennett1993}, $\textit{We}$ needs to be replaced with an effective Weber number ${\textit{We}}^{\dagger}=D_0/(D_0+2Z_c^*)\textit{We}$. Furthermore, during the impact the droplet experiences an average deceleration force with magnitude $\sim E_d/\tfrac{1}{2}D_0$ [Eq.~\ref{eq.energy_ratio}]. {Reproducing} %By analogy of 
the argument in the force balance model introduced in ref. \citenum{CLANET2004}, the droplet deforms to balance such a deceleration, which yields {a} %the 
scaling with ${\textit{We}}^{\dagger}$ as well. Therefore, we suggest to use ${{\textit{We}}^{\dagger}}=\frac{D_0}{D_0+2Z_c^*}\textit{We}$  for the impact on a deformable substrate, where $Z_c^*$ characterizes the deformation of the substrate.

{When t}he maximum droplet deformations are plotted versus effective Weber {number} in Fig.~\ref{f.dmax}b{, we find that the d}ata collapse {onto} a master curve. A power law fit gives {${{{\textit{We}}}^{\dagger}}^{0.250\pm0.012}$, } %${{{\textit{We}}}^{\dagger}}^{0.2495\pm0.012}$. 
{which implies that our experiments are well described by}
the force balance model {of} {Clanet \textit{et al}}\cite{CLANET2004}. Previous studies on this topic {rather} used {the traditional Weber number} %ordinary $\textit{We}$ for the scaling 
and reported smaller scaling exponents no larger than $1/5$\cite{Marston2010,Nefzaoui2012}. {This had} %It has 
been interpreted either as {a viscous} effect\cite{Nefzaoui2012} or {as a result of the} density {ratio} of the liquid {and sand}\cite{Marston2010}.
However, liquids with the same viscosity and density but different surface tension scale differently\cite{Nefzaoui2012}. We speculate that, other than viscosity, the impact energy distribution in Eq.~\ref{eq.energy_ratio} may be helpful to understand those inconsistent observations {in the literature}. The mechanism leading to Eq.~\ref{eq.energy_ratio} is that the impact energy $E_k$ is distributed according to the relative stiffness of the intruder and the target, \textit{i.e.}, {the balance goes always towards the most easy deformation.} %direction. 
The extreme case is that either the intruder or the target is undeformable, {in which case the energy} quickly transfers into the {deformable medium}. When both intruder and target are deformable, the impact energy dissipates more into the `softer' one. For instance, while impacting on the same substrate, a liquid droplet with a smaller surface tension is easier to deform, {and} in consequence the substrate behaves more solid-like, and $\textit{We}^{1/4}$ is more likely {to be} recovered. This %annotates 
{could explain} the different scaling {exponents found for liquids that} only {differ in} surface tension in ref. \citenum{Nefzaoui2012}. The computation of Eq.~\ref{eq.energy_ratio} requires the maximum crater depth, $Z_c^*$, which is measured {for the} first time in the {present} experiments.

\section{Liquid-grains mixture and crater morphology}
%\label{sec.mixing}

Until so far we %demonstrate 
{considered} the deformation of the target and that of the intruder separately. However, they are `miscible' as well, \textit{i.e.}, the liquid {that composes the} droplet penetrates into the granular substrate. {This} penetration results in a liquid-grain mixture which {needs to be taken into account in order} %is important 
to understand the crater morphology\cite{Delon2011,Nefzaoui2012,Zhao2014}. By changing impact speed $U_0$ and the initial packing fraction $\phi_0$ the crater morphology is found to vary systematically. According to the residue shape the observed craters are categorized into three groups: doughnut, truffle, and {pancake} [Fig.~\ref{f.morph}]. The residue consists of a mixture of liquid and grains, which is referred to as `liquid marble' in ref. \citenum{Zhao2014}. The discrimination between doughnut and truffle is vague{, as while changing the impact parameters one shape seems to be continuously transformed into the other, whereas} the transition from truffle to pancake is more abrupt. In this section, a model about the liquid penetration is established together with the knowledge already obtained in the previous {s}ections. %above. 
We will use this model to explain {the} various crater morphologies.

The doughnut and truffle residues are formed from the grains mixed with the liquid which are {transported} %taken 
towards crater center by the droplet contraction. If the penetration during the impact is little, after the contraction, pure liquid concentrates at the crater center surrounded by the mixture of liquid and grains. After a few {hundreds of milliseconds}, the liquid in the center penetrates into the substrate due to gravity and capillary force leaving a dimple, \textit{i.e.}, {creating} a doughnut residue. More penetration results in {a smaller amount of} %less 
pure liquid at the center, {such that} %thus 
the residue gradually becomes a truffle. With even more penetration the droplet {is hardly able to contract}, leaving a flat residue, {the} pancake shape. Therefore, {knowing} the amount of penetration of liquid between grains is %prominent 
{crucial} to understand the morphology phase diagram. %[Fig.~\ref{f.setup}e].

\begin{figure}
\centering
  \includegraphics[width=8.3cm]{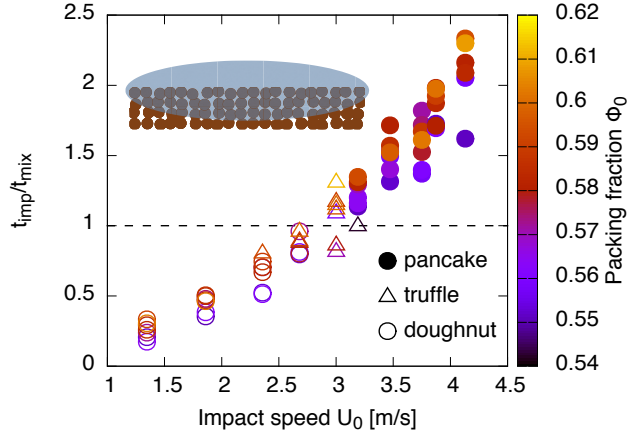}
  \caption{\label{f.phase} Phase diagram of the crater morphology. {On the y-axis we plot} the ratio between two time scales, {namely the} impact time scale $t_{imp}$ and {the} mixing time $t_{mix}$. The {horizontal} dashed line indicates {a} ratio {of} 1, {where the transition is expected. On the x-axis we have the impact speed $U_0$.} The various crater morphologies are labeled with different symbols {(see legend), whereas the color of the symbols indicate} initial packing fraction. {The inset shows a schematic of the droplet for which $V=V^*$ at maximum expansion.}}
\end{figure}

The transition from doughnut to truffle is continuous, while a sharper transition between truffle and pancake is observed. We %use 
{estimate} the volume of the liquid-grain mixture to characterize the penetration amount and define a threshold volume $V^*$ to separate doughnut/truffle and pancake regime. When this threshold mixture volume, $V^*$, is reached, the droplet loses the high curvature {edge that promotes} %leading 
contraction{. This happens when} %, \textit{i.e.}, 
the volume of the mixture in the substrate, $V^*$, equals the volume of pure liquid above the substrate, $V_l-V^*(1-\phi)$, where $V_l$ is the initial droplet volume [Fig.~\ref{f.phase} inset]:
\begin{equation}
V^*=\frac{V_l}{2-\phi}.
\label{eq.critical_vol}
\end{equation}
If the time scale to reach this critical mixing volume, $t_{mix}$, is shorter than the impact time scale $t_{imp} = (D_0+2Z_c^*)/U_0$, the droplet contraction is suppressed, and pancake shapes are observed. Otherwise, the droplet {is able to contract}, mixed with grains, and forms a doughnut/truffle residue. % The reason choosing $t_{imp}$ as the reference time scale is explained with hindsight.

To quantitatively examine the above picture, we need to formulate the mixing progress. Describing the granular substrate as a porous medium, we start with Darcy's law:
\begin{equation}
\frac{Q}{A}\approx\frac{\kappa P}{\mu L}.
\label{eq.darcy}
\end{equation}
Here, the permeability of the granular packings, $\kappa$, is defined by Carman-Kozeny relation {$\kappa = (1-\phi)^3 d_g^2/(180 \phi^2)$, where} $Q$ is the volume flux into the porous substrate, $P$ the driving pressure,  $\mu$ the {dynamic} viscosity of the liquid, $L$ the penetration depth of the liquid into the porous substrate, and $A$ the contact area. {In Eq.~}\ref{eq.darcy}{ we estimate the pressure gradient inside the sand as $P/L$.} Furthermore, the conservation {of liquid volume} calls for
\begin{equation}
\frac{Q}{A} = (1-\phi)\frac{dL}{dt}.
\label{eq.vol_con}
\end{equation}
Here, we only consider $L$ as a function of time. The total liquid volume penetrating into the substrate is given by $(1-\phi)AL$.

Combining Eq.~\ref{eq.darcy} and Eq.~\ref{eq.vol_con}, the %dynamic of the 
penetration depth $L$ is solved as $L(t) = \sqrt{2P\kappa/(\mu(1-\phi))\,t}$. From here, one can define the time scale for which the volume of the mixture reaches the critical volume, $AL(t_{mix})=V^*$:
\begin{equation}
t_{mix} = \left( \frac{V^*}{A} \right)^2 \frac{(1-\phi)\mu}{2\kappa P}.
\label{eq.tmix}
\end{equation}

To apply Eq.~\ref{eq.tmix} a few quantities need to be evaluated and explained. The contact area is estimated as $A=\pi {D_d^*}^2/4${, \textit{i.e.}, the contact area at the measured maximum spreading diameter}. The packing fraction $\phi$ is evaluated by the critical packing fraction $\phi^*$%{defined in subsection~}\ref{subsec.speed} 
rather than the initial packing fraction $\phi_0$ as explained in Fig.~\ref{f.dilation_mech}. The last missing piece is the driving pressure $P$. There are three candidates: the gravitational pressure, the inertial pressure, and the capillary pressure {caused by the hydrophilicity of the grains that tries to pull the liquid into the bed}. Since the droplet diameter is at the magnitude of the capillary length, gravity cannot {significantly} deform the droplet. The large droplet deformation shown in Fig.~\ref{f.dmax} indicates that the inertial pressure is much larger than gravity. A simple experiment of zero impact speed{, in which we determine how fast the droplet is absorbed by the bed,} also indicates that for the used grains capillary pressure is at least two order of magnitude smaller than the inertial pressure. Therefore {it is justified to} consider the inertial pressure as the driving pressure. In the analysis of maximum droplet deformation {and} the introduction of the effective Weber number ${\textit{We}}^{\dagger}$ {it was already implied that %already implies 
the} average deceleration force experienced by the droplet {equals} $E_d/\tfrac{1}{2}D_0$ [Eq.~\ref{eq.energy_ratio}]. The inertial pressure is evaluated in consequence as: 
\begin{equation}
P = \frac{E_d}{\frac{1}{2}D_0 A}=\frac{M_lU_0^2}{A(D_0+2Z_c^*)}.
\label{eq.pressure}
\end{equation}
With all the quantities in Eq.~\ref{eq.tmix} {now defined}, we can finally compare the time scales, $t_{imp}$ and $t_{mix}$, in Fig.~\ref{f.phase}.

Three features from the experimental phase diagram are recovered in Fig.~\ref{f.phase}: (i) the transition is dominated by impact speed; (ii) a denser packing induces more mixture; {and} (iii) the expected transition discriminating doughnut/truffle and pancake morphology around $U_0\approx 3.2\, \mathrm{m/s}$ is {indeed} indicated by {the condition} $t_{imp}=t_{mix}$. We {emphasize} that the inertial pressure driving penetration during the droplet spreading is considered to {dominate} the crater morphology transition, rather than gravity during the recession as {was suggested} in ref. \citenum{Delon2011}. Therefore, the reference time scale is the characteristic period of the inertial pressure, $t_{imp}$, rather than the contact time scale $\sim\sqrt{\rho_l D_0^2/\sigma}$\cite{Zhao2014}. Finally, it is worthy to note that according to Eq.~\ref{eq.tmix} and Eq.~\ref{eq.pressure} the ratio of the time scales $t_{imp}/t_{mix}$ is independent of the substrate deformation $Z_c^*$. To apply and check the model, only $D_d^*$ needs to be measured.

\section{Conclusion}
In this paper, we study the deformation of the droplet and the granular packing during impact. The deformation of the granular substrate decreases with the initial packing fraction [Fig.~\ref{f.zmax}], however, the deformation speed exhibits evidence of dilation and defines a critical packing fraction $\phi^*\approx 0.585$ in Fig.~\ref{f.dilation_mech}. The substrate deformation introduces an impact energy distribution, Eq.~\ref{eq.energy_ratio}, between the intruder and the target. An effective Weber number, ${\textit{We}}^{\dagger}$, is defined accordingly, and the droplet maximum deformation, $D_d^*$, {is shown to be consistent with a scaling law $D_d^* \sim {{\textit{We}}^{\dagger}}^{1/4}$} in Fig.~\ref{f.dmax}, which suggests a scenario where surface tension balances{ }%the 
inertial pressure. Finally, based on the results of $\phi^*$ and $D_d^*$, a model is established to describe the penetration of the liquid into the substrate. This model evaluates the competition between the penetration time and the impact time, {from which it is able to explain} the observed morphology transition {between the doughnut/truffle and the pancake crater shapes} [Fig.~\ref{f.phase}]. The energy distribution as given in Eq.~\ref{eq.energy_ratio}, which is essential to understand the deformation of the target and the intruder, is estimated with the average interaction force. Its validation needs time resolved measurement of the interaction force.

We acknowledge the discussion with X. Cheng. This project is supported by {FOM and NWO through a VIDI Grant No. 68047512.}

\bibliography{drop_impact} 

\begin{thebibliography}{10}
\bibitem{note1}
When the droplet reaches its maximum spreading diameter, its average thickness
  $h = (2/3) D_0 (D_0/D_{max})^2$. According to the geometric configuration of
  the experiments, the depth difference introduced by this thickness is ~
  $0.37h$. For the smallest $U_0$, the least spreading case, where the
  refraction introduces the largest error, this estimation gives a height
  difference of 0.14 mm, which is close to the depth resolution of the laser
  profilometry system. Moreover, this refraction error is well below the
  smallest measured $Z_c^*$ of 0.5 mm. Since the center of the impact droplet
  is even thinner than $h$ as estimated above, the influence on the maximum
  center depth measurement would be even weaker in practice.
\end{thebibliography}
\bibliographystyle{unsrt}

\end{document}